\font\manual=manfnt

\def\Q{\mathbf{ Q}}
\def\R{\mathbb{ R}}

\def\LL{{\cal L}}

\def\RootOf{\mathop{\rm RootOf}\nolimits}

\documentclass{llncs}
\usepackage[hyphens]{url}
\usepackage{pdfpages}
\usepackage{verbatim}
\usepackage{hyperref}
\usepackage{combelow}
\usepackage[show]{ed}
\usepackage{graphicx}
\usepackage{amsmath}
\usepackage{amsfonts}
\usepackage{lscape}
\newtheorem{notation}{Notation}

\begin{document}

\author{Erika {\'A}brah{\'a}m\inst{1} \and James Davenport\inst{2} \and Matthew England\inst{3} \and Gereon Kremer\inst{1}\and Zak Tonks\inst{2}}

\institute{RWTH Aachen University, Germany\\ \email{\{abraham,gereon.kremer\}@cs.rwth-aachen.de}\and
  University of Bath, Bath BA2 7AY, UK\\
  \email{\{J.H.Davenport,Z.P.Tonks\}@bath.ac.uk}
\and Coventry University, UK\\ \email{Matthew.England@coventry.ac.uk}
}

\title{New Opportunities for the Formal Proof of Computational Real Geometry?}

\maketitle

\begin{abstract}
The purpose of this paper is to explore the question ``to what extent could we produce formal, machine-verifiable, proofs in real algebraic geometry?'' The question has been asked before but as yet the leading algorithms for answering such questions have not been formalised.  We present a thesis that a new algorithm for ascertaining satisfiability of formulae over the reals via Cylindrical Algebraic Coverings [\'{A}brah\'{a}m, Davenport, England, Kremer, \emph{Deciding the Consistency of Non-Linear Real Arithmetic Constraints with a Conflict Driver Search Using Cylindrical Algebraic Coverings}, 2020] might provide trace and outputs that allow the results to be more susceptible to machine verification than those of competing algorithms.
\end{abstract}


\section{Setting}

Computational Real Algebraic Geometry really began with the Cylindrical Algebraic Decomposition (CAD) work of Collins \cite{Collins1975}, and independently W\"uthrich \cite{Wuthrich1976}: previous results such as \cite{Seidenberg1954,Tarski1951} having been effective in name only.

\begin{definition}
An {\em algebraic proposition\/} is one built up from expressions of the form $p_i(x_1,\ldots,x_n)=0$ (where the $p_i$ are polynomials with integer coefficients) joined together by the logical connectives $\lnot$ (not), $\land$ (and) and $\lor$ (or). A {\em semi-algebraic proposition\/} is the same, except that the building blocks are expressions of the form $p_i(x_1,\ldots,x_n)\sigma0$ where $\sigma \in \{=,\ne,>,\ge,<,\le\}$. The language of semi-algebraic propositions is also called the \emph{Tarski language} $\LL$.
\end{definition}

\begin{notation}
\label{N1}
Although not strictly in the Tarski language, we will find it convenient to describe ``the $i$-th root of $p$'' by ${\strut}_i\sqrt[n]{p}$ (or more generally ${\strut}_i\RootOf(p,y)$) to mean the $i$th real root (counting from $-\infty$) of $p$, as a polynomial in $y$ ($\strut_ip$ when the variable is clear).
\end{notation}
Thom's Lemma \cite{CosteRoy1988} means that these can be converted into statements in $\LL$, at the cost of adding additional polynomials --- the derivatives of the ones we have.

\begin{notation}
We let $\R^*$ be $\R$ to which we add $\epsilon$ to make the ordered ring $\R[\epsilon]$ with $0<\epsilon<$ any positive real, and then add $\pm\infty$ to the underlying ordered set.
\end{notation}
See \cite{Kosta2016a} for the rationale behind these symbols in Virtual Term Substitution.

\begin{problem}[Quantifier Elimination]\label{Pr:1}
Consider a quantified proposition\footnote{Any proposition with quantified variables can be converted into one in this so-called
{\em prenex normal form\/} 
 --- see any standard logic text.}
\begin{equation}\label{eq:QE}
Q_1 y_1\ldots Q_m y_m F(y_1,\ldots,y_m,x_1,\ldots,x_n),
\end{equation}
where $F\in\LL$ and $Q_i \in \{\exists,\forall\}$.  Does there exist a quantifier-free equivalent semi-algebraic proposition $G(x_1,\ldots,x_n)$ and if so, can we compute it?
\end{problem}

\begin{notation}
The structure of (\ref{eq:QE}) forces an order onto the variables, which we take from first to last as $x_1,\ldots,\allowbreak x_n,y_1,\ldots,y_m$.
\end{notation}
The most used complete method to solve Problem \ref{Pr:1} is Cylindrical Algebraic Decomposition (CAD) \cite[and much subsequent work]{Collins1975}, although Virtual Term Substitution (VTS) \cite[and much subsequent work]{Weispfenning1988} is very useful when it is applicable: see, for example, \cite{Tonks2020a}.  
Problem \ref{Pr:1} is known to be doubly-exponential (in $n+m$) in the worst case \cite{BrownDavenport2007,DavenportHeintz1988}, but more accurately it is doubly-exponential in the number of times the sequence of $Q_i$ changes from $\exists$ to $\forall$ or \emph{vice versa} \cite{Basu1999}.

An important special case of Problem \ref{Pr:2} is the following.
\begin{problem}[Satisfiability]\label{Pr:2}
Given a fully existentially quantified proposition
\begin{equation}\label{eq:P2}
\exists x_1\exists x_2\cdots\exists x_n F(x_1,\ldots,x_n),
\end{equation}
where $F\in\LL$, does there exist a solution?  I.e. is this {\bf true} or {\bf false} (also known as the problem being {\bf SAT} or {\bf UNSAT})?
\end{problem}
Despite the fact that (\ref{eq:P2}) is a quantified piece of algebra, the problem is known as ``Quantifier-Free Non-linear Real Arithmetic'' (\verb!QF_NRA!) in the Satisfiability Modulo Theories (SMT) community \cite{Barrettetal2017a}, since all variables in SMT are assumed to be existentially quantified anyway. Traditionally Problem \ref{Pr:1} was the focus of the symbolic computation community and Problem \ref{Pr:2} the focus of the Satisfiability Checking community, but recently there has been a joint effort \cite{SC2}.

By \cite{Basu1999} it is soluble in time singly-exponential in $n$, but the authors know of no implementation of this.
A fundamental difference between Problem \ref{Pr:2} and the famous Boolean version of Satisfiability Checking is that there are an infinite number of possible values for the $x_i$, so direct `brute force' is not an option.

A subtle variant of Problem \ref{Pr:2} is the following.
\begin{problem}[Proven Satisfiability]\label{Pr:3}
Given a fully existentially quantified proposition (\ref{eq:P2}),
where $F\in\LL$, produce a computer-verifiable proof of {\bf SAT} or {\bf UNSAT}.
\end{problem}
If the answer is {\bf SAT}, all the major algorithms (at least VTS and CAD, as well as hybrids of these methods) will actually compute the witnesses $x_i$, so the computer-verifiable proof of {\bf SAT} is relatively easy (we say relatively easy as we must bear in mind that these $x_i$ might be algebraic numbers). The challenge is the {\bf UNSAT} case.  The main body of work on this problem is by Cohen and Mahboubi \cite{Mahboubi2006}, \cite{Mahboubi2007}, \cite{CM10}, \cite{CohenMahboubi2012a}.  In the former works an attempt was made to formalise CAD but to the best of our knowledge this was not completed. In those latter papers QE is verified but not via CAD but with an algorithm described in \cite{Hormander1983}, which falls in the ``effective in name only'' category.

\subsection*{Thesis and plan of this article}

The authors of the present paper have recently developed a new algorithm for tackling Problem \ref{Pr:2} in \cite{Abrahametal2020a} which offers computational advantages over CAD.  The algorithm is based around the new idea of Cylindrical Algebraic Coverings and so we refer to it here as CAC.  The present article is essentially a position paper where we present our thesis that the UNSAT results produced by CAC may be far more susceptible to formal proof than those of the traditional tools. We present examples which show that CAC produced proofs that are much closer to those of a human.  The aim of the paper is to present this thesis to the formal proof community to garner interest and insight into whether it may be followed.

The paper continues by introducing the two leading technologies for tackling our problems in Section \ref{SEC:Background}.  Then in Section \ref{SEC:CAC} we summarise our recent CAC algorithm.  In Section \ref{SEC:Examples} we present some examples that illustrate our thesis.

\section{Traditional algorithms for these problems}
\label{SEC:Background}

\subsection{Cylindrical Algebraic Decomposition}

This was first introduced by Collins in \cite{Collins1975}. The key idea is to partition $\R^n$ into connected subsets called cells with the following properties:
\begin{enumerate}
\item Each cell is sign-invariant for all the polynomials in $F$.
\item Each cell $C$ has a sample point $s_C$ identified within it.
\item The cells and sample points are arranged cylindrically.  This means that for all $k<m+n$, the projections onto the first $k$ variables of two cells are either equal or disjoint, and if equal their sample points will have the same values for the first $k$ coordinates.
\item The cells have semi-algebraic descriptions.
\end{enumerate}
Then the truth of $\exists y F(y,x_1,\ldots,x_n),$ at a sample point $s=(s_1,\ldots,s_n)$ follows from the truth of \emph{any} of the sample points $s'=(s_1,\ldots,s_n,s_{n+1})$ above it (and if they are all false, then $\exists y F$ is false). Similarly, the truth of $\forall y F(y,x_1,\ldots,x_n),$ at $s$ follows from the truth of \emph{all} of the sample points $s'$ above it (and if any of them are false, then $\forall y F$ is false).
The truth of $\exists y$ from the truth at any sample point (and the falsity of $\forall y$ from the falsity at any sample point) are obvious: the converses follow from the completeness of the decomposition, i.e. that there is no behaviour not captured by a sample point.

There are improvements to the original \cite{Collins1975} by many authors: in particular \cite{McCallum1984} has a more efficient computation, but may \emph{explicitly} state that the decomposition is not complete (``nullification'' or ``not well ordered''), and \cite{Lazard1994} (justified only recently in \cite{McCallumetal2019a}), is more efficient still but always complete. Further efficiencies can be found when $F$ has certain structure, for example, \cite{McCallum1999a} considered the case where $F$ has the form $p(y_1,\ldots,y_m,x_1,\ldots,x_n)=0 \land F'(y_1,\ldots,y_m,x_1,\ldots,x_n)$ where $p$ is referred to as an \emph{equational constraint}; \cite{McCallum2001}, \cite{Englandetal2019a} considered the case of several equational constraints; and \cite{Bradfordetal2016a} considered more complicated combinations, re-defining (1) in the construction to:
\begin{enumerate}
\item[1$'$] Each cell is truth-invariant for $F$, but not necessarily sign-invariant for all the polynomials in $F$.
\end{enumerate}
See for example the introduction of \cite{Englandetal2019a} for a more detailed review.  The key point is that however the decomposition is constructed, the fundamental requirement is \emph{completeness}, i.e. that all cases, however defined, are captured as a cell (sample point) in the decomposition.

\subsection{Virtual Term Substitution}

Virtual term Substitution (VTS) was introduced in \cite{Weispfenning1988}, and many developments are gathered in \cite{Kosta2016a,Kostaetal2016a}. Here the key idea for $Q y F(y,x_1,\ldots,x_n)$ is to consider a $y$-value from every interval of the real line according to those intervals formed by the real roots of all polynomials contained in $F$, regarded as elements of $\Q[x_1,\ldots,x_n][y]$. Unlike CAD, it is limited to elimination of quantified variables appearing as low degree - currently the methodology is described for up to degree 3 in \cite{Kosta2016a}. Action of elimination by VTS may even increase the degree of intermediate formulae, and as such all $x_1,\dots,x_n$ appearing at most cubically is not a guarantee VTS can complete QE alone.

If $n=0$, these would truly be values $v_i$ in $\R^*$, the truth of $\exists y F(y)$ would be that of $\bigvee_i F(v_i)$, and the truth of $\forall y F(y)$ would be that of $\bigwedge_i F(v_i)$.
In general, of course, $n>0$, so we have terms rather than values, which are the roots in $y$ of these polynomials in $\Q[x_1,\ldots,x_n][y]$, and the substitution into $F$ is ``virtual''.  We write  $F[y/\hskip-2pt/v_i]$ for ``the virtual substitution of $v_i$ for $y$ in $F$'', which is especially relevant if the $v_i\in\R^*\setminus\R$.

Again, if a $F(y/\hskip-2pt/v_i)$ is true, then $\exists y F(y)$ is true, and if a $F(y/\hskip-2pt/v_i)$ is false, then $\forall y F(y)$ is false.  Deducing the converse again requires that the set of $v_i$ be ``complete''. If $v_i\in\R^*\setminus\R$ is a witness to truth/falsity, then it is possible, with the hindsight of knowing the expressions in which it occurs, to replace $\infty$ by a large enough number, and $\epsilon$ by a small enough one - see Algorithm 1 from \cite{Tonks2020a}.

\section{Cylindrical Algebraic Coverings}
\label{SEC:CAC}

We recently presented a new algorithm for determining the satisfiability of conjunctions of non-linear polynomial constraints over the reals \cite{Abrahametal2020a}, which can be used to solve Problem \ref{Pr:2}. The algorithm is based around the technology of CAD but does not build a decomposition of $\R^n$. Instead, overlapping cells are generated, until we have a covering of the sample space.  Sample points are constructed incrementally, either until a satisfying sample is found or sufficient samples have been sampled to conclude unsatisfiability.  The choice of samples is guided by both the input constraints and previous conflicts (combinations of constraints and samples found to be unsatisfiable).  The key idea behind our new approach is to start with a partial sample; demonstrate that it cannot be extended to a full sample; and from the reasons for that rule out a larger space around the partial sample, which build up incrementally into a covering of the space.  The cells are still arranged in cylinders and have semi-algebraic descriptions and thus we call the data structure produced a Cylindrical Algebraic Covering (CAC).

Unlike CAD, which starts with projection to generate algebraic information, the algorithm described in \cite{Abrahametal2020a} starts with ``guessing'' a sample point dimension-wise, starting in the lowest dimension and iteratively extending it to higher dimensions. Either we ``guessed'' right and find a satisfying sample or we face a partial sample that cannot be extended to a full solution and use it to guide the projection (and thus the cell construction). To do so, we recursively execute the following:
\begin{itemize}

\item Given an $i$-dimensional sample $s=(s_1,\ldots,s_{i})$, we try to extend it to a sample $(s_1,\ldots,s_i,s_{i{+}1})$  that does not evaluate any input constraint to false.

\item If this works out then either the sample is full dimensional and we report consistency (and this witness), or we continue with extending the sample in the next dimension.

\item Otherwise we take note of the reason the sample cannot be extended, and exclude from further search not just this particular sample $s$, but all extensions of $(s_1,\ldots,s_{i{-}1})$ into the $i$th dimension with any value from a (hopefully large) interval around $s_i$ which is non-extensible for the same reason.

\item We continue and check further extensions of $(s_1,\ldots,s_{i{-}1})$ until either we find a solution or the $i$th dimension is fully covered by excluding intervals.

\item In the latter case, we analyse the collection of intervals to try and rule out not just the original sample in $\mathbb{R}^{i{-}1}$ but an interval around it within dimension $(i{-}1)$, i.e. the same procedure we benefited from in the recursive call.

\end{itemize}
The generalisation of the unsatisfying sample to a wider interval is based upon CAD technology, i.e. we know the truth of a constraint can only change when we cross the real roots of certain projection polynomials calculated with tools such as coefficients, discriminants and resultants.  Intuitively, each sample $s\times s_i$ violating a constraint with polynomial $p$ can be generalised to a cell in a $p$-sign-invariant CAD. So when all extensions of  $s$ have been excluded (the $i$th dimension is fully covered by excluding intervals) then we project all the covering cells to dimension $i{-}1$ and exclude their intersection from further search.

The conflict generalisation guides the algorithm to sample away from the reasons of previous conflicts, which should allow for a satisfying sample to be found quicker if one exists.  In the case of UNSAT, a covering in which every cell is UNSAT could use less cells than an entire decomposition.

See  \cite{Abrahametal2020a} for full details of the algorithm including worked examples and details of experimental results.
The CAC based algorithm has similarities with the incremental variant of CAD, the NLSAT method of Jovanovi\'{c} and de~Moura \cite{JdM12}, and the NuCAD algorithm of Brown \cite{Brown2015} but the examples in \cite{Abrahametal2020a} demonstrate its unique advantages.

In the present paper we hope to demonstrate another potential advantage:  the increased susceptibility to formal verification of its output.

\section{Examples}
\label{SEC:Examples}

We demonstrate our thesis with some simple UNSAT examples.

\begin{figure}
\centering
\includegraphics[width=0.45\textwidth]{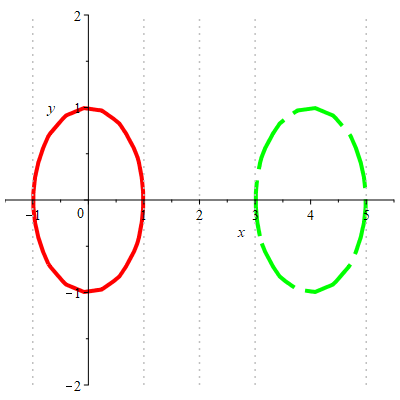} \qquad 
\includegraphics[width=0.45\textwidth]{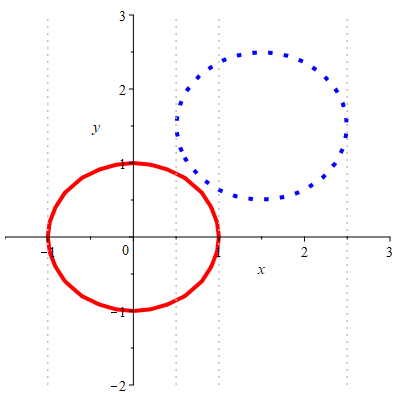}
\caption{Graphs of polynomials involved in the worked examples.}
\label{pic}
\end{figure}

\subsection{Example 1}
\label{Ex:1}

Consider $F:=(x^2+y^2<1)\land((x-4)^2+y^2<1)$. The circles are graphed on the left of Figure \ref{pic} and we see they do not intersect and thus $F$ must be UNSAT. 

\subsubsection{CAD:}

To solve Problem \ref{Pr:2}, the CAD algorithm would do the following:
\begin{itemize}
\item[(a)] partitions the $x$-axis at\footnote{$x=\pm1$ are the extremal points of the first circle, and $x=3,5$ of the second. One could ask why $x=2$. The answer is that both circles (boundaries of the discs) in $F$ have common zeros at $x-2,y=\pm\sqrt{-3}$: of course these zeros are not real, but have a real $x$-component.} $-1,1,2,3,5$;
\item[(b)] constructs 27 cells and associated sample points in $R^2$ (see Table \ref{T:Ex1});
\item[(c)] deduces that no sample points have both $x^2+y^2<1$ and $(x-4)^2+y^2<1$ true at once; 
\item[(d)] and therefore (since the polynomials are sign-invariant in the cells) concludes that no cells have both true anywhere over them;
\item[(e)] and because the union of cells is $\R^2$, the statement must be nowhere true.
\end{itemize}

Step (c) is analogous to verifying {\bf SAT}, and, due to the cylindrical nature of the decomposition, (e) is relatively easy. The problem is verifying (d). Its truth depends on the fine details of the CAD algorithm used (for example \cite{McCallum1984} is distinctly different from \cite{Collins1975}, and the Lazard method \cite{McCallumetal2019a} is based on entirely different mathematics again).

\subsubsection{VTS:}

Virtual Term Substitution would consider a variety of possible values $v_i$ for $y$. In the implementation of \cite{Tonks2020a}, the set of $v_i$ starts with $-\infty$, then various other trivial (in the sense of immediately yielding false) values, and the first non-trivial one is ${\strut}_1\sqrt{1-x^2}+\epsilon$. Virtually substituting this into $F$ yields 
\begin{equation}
\underbrace{x^2<1}_{\hbox{guard}}\land \underbrace{\hbox{true}}_{x^2+y^2<1}\land \underbrace{\left(-x<-2\lor\left(x=2\land x^2<1\right)\right)}_{(x-4)^2+y^2<1},
\end{equation}
where the guard is there to make sure that the substitution makes sense, and ``$-x<-2$'' is the simplification of $(x-4)^2+\left({\strut}_1\sqrt{1-x^2}+\epsilon\right)^2<1$. In all there are 41 VTS test points, of which 21 are initial ones used in $y$, and of the remaining 20 on $x$, there are 7 distinct ones (used on similar intermediate formulae). They are characterised in a manner intelligible as semi-algebraic sets (to compare with CAD) in Table \ref{T:Ex1VTS}. It is important to note that this is only a characterisation, considering VTS does not in itself operate geometrically.  We have fewer substitutions of exact values for $x$ (what would be analogous to computing ``sections'' in CAD). 

Speaking generally, VTS is an algebraic approach on formulae as opposed to geometry.  Amongst the formulae produced in $x$, $x=2$ appears, but other atomic formulae are all strong relations, such that this is the only substitution of an ``exact'' value. All of the generated test points are substituted in order to deduce {\bf UNSAT}, as we form a disjunction of formulae all equivalent to false.

\begin{table}
\caption{Structural Test Points for VTS (as semi-algebraic sets) for the example in Section \ref{Ex:1}\label{T:Ex1VTS}}
\center \begin{tabular}{c|c|c|c|c|c|c}
$x<-1$ \,&\, $-1<x<1$ \,&\, $1<x<2$ \,&\, $x=2$ \,&\, $2<x<3$ \,&\, $3<x<5$ \,&\, $x>5$
\end{tabular}
\end{table}

\subsubsection{Human:}

Of course no human prover would proceed in either of these ways. A human produced argument would be along the lines of 
\begin{align*}
x^2+y^2<1 &\Rightarrow x^2<1 \Rightarrow x<1\\
(x-4)^2+y^2<1 &\Rightarrow (x-4)^2<1 \Rightarrow x-4 \in (-1,1) \Rightarrow x-4>-1 \Rightarrow x>3 
\end{align*}
with the two right most statements clearly incompatible.

\subsubsection{CAC:}

Cylindrical Algebraic Covering proceeds by choosing a variety of sample $x$ values, then recursing on $y$ (and any subsequent variables if there were any. In this example it it proceeds as follows, but we note that the theoretical algorithm allows a great deal of choice in computation path.
\begin{description}
\item[$x=-1$:] This would require $y^2<0$ which is unsatisfiable. However, we can deduce nothing about the neighbouring values of $x$, as $-1$ is a root of the discriminant of $x^2+y^2-1$.
\item[$x<-1$:] We sample $x=-2$ and find this is also impossible for the same reason.  The nearest root of this resultant is $-1$, so $(-\infty,-1)$ is ruled out along with $x=-2$.
\item[$x>-1$:] We sample $x=0$.  Here $y\in(-1,1)$ can not be ruled out by $x^2+y^2<1$ (obviously), but $(x-4)^2+y^2<1$ rules out this value of $x$ immediately, and the generalisation rules out with it the whole of $(-\infty,3)$.
\item[$x\ge3$:] We sample $x=4$.  This trivially conflicts with $x^2+y^2<1$, which rules out $(1,\infty)$.
\item[\rm{Hence}] the whole of $\R$ is ruled out for $x$, and we may conclude UNSAT.
\end{description}
While this is not quite as simple as the human proof above, it is much closer to it. If we pruned the reasoning, it would be that $x\in(-\infty,3)$ is infeasible because of $(x-4)^2+y^2<1$, and $x\in(1,\infty)$ because of $x^2+y^2<1$.  Most importantly, the argument of unsatisfiability can be reconstructed from the algorithm flow and output.

\subsection{Example 2}
\label{Ex:2}

Consider $F:=(x^2+y^2<1)\land\left(\left(x-\frac32\right)^2+\left(y-\frac32\right)^2<1\right)$.    The circles are graphed on the right of Figure \ref{pic} and we see they again do not intersect and thus $F$ must be UNSAT.  

\subsubsection{CAD:}

To solve Problem \ref{Pr:2}, the CAD algorithm behaves similarly to Example 1: the two circles have critical points (roots of the discriminant) at $x=-1,x=1$ and $x=\frac12,x=\frac52$ respectively. This time the resultant of the two circles is $18x^2-27x+\frac{45}4$, whose roots are not real numbers and so do not contribute to the set of critical points.  However, because the circles overlap the cylinders above the $x$-axis are more decomposed and we have 41 cells as in Table \ref{T:Ex2}.

\subsubsection{VTS:}

VTS this time receives non-false formulae owing to substitutions of test points in $y$ from both circles. Virtual substitution can substitute roots directly from multivariate polynomials (which, in fact, is some of the cadence behind the terminology ``virtual''). An example is the virtual substitution of $\strut_1\RootOf{}\left( \left(x-\frac32\right)^2+\left(y-\frac32\right)^2-1 \right) + \epsilon$ for $y$ into $F$, leading to (after some minor simplification):
\begin{align*} 
4x^2 - 12x + 5 < 0 \land \Bigl( & x<0  \land -8x^2 + 12x - 5<0 \,\lor\, \\ & 8x^2 - 12x + 5 < 0 \,\lor\, \\ & x \geq 0 \land 8x^2 - 12x + 5 = 0 \land 2x^2 - 6x + 7 < 0 \Bigr) 
\end{align*}
Now, $4x^2 - 12x + 5 = (2x-1)(2x-5)$, but $8x^2 - 12x + 5$ and $2x^2 - 6x + 7$ are irreducible. As a result of the irreducible polynomials, in contrast to Problem \ref{Pr:1}, VTS must then use test points from quadratic polynomials in $x$ rather than using purely linear test points. In fact, $8x^2 - 12x + 5$ has no real roots, and the negative discriminant manifests in a guard evaluating as false to prevent substitution from this. In total, in the same manner as Table \ref{T:Ex1VTS} we have Table \ref{T:Ex2VTS} as the characterisation of the geometry in $x$ that VTS finds relevant.

\begin{table}
\caption{Structural Test Points for VTS (as semi-algebraic sets) for the example in Section \ref{Ex:2}\label{T:Ex2VTS}}
\begin{tabular}{lc|cl}
$x<\strut_1\RootOf{}(2x^2 + 6x - 7)$ 
& \hspace{0.2in} & \hspace{0.2in} &$x=\strut_1\RootOf{}(2x^2 - 6x + 7)$ 
\\ \\
$\strut_1\RootOf{}(2x^2 - 6x + 7)<x<0$ 
& \qquad && $x=0$ 
\\ \\
$0<x<\frac12$ 
& \qquad && $\frac12<x<\strut_2\RootOf{}(2x^2 - 6x + 7)$ 
\\  \\
$x=\strut_2\RootOf{}(2x^2 + 6x - 7)$ 
& \qquad && $\strut_2\RootOf{}(2x^2 - 6x + 7)<x<\frac52$ 
\\ \\
$x>\frac52$ & \qquad && 
\end{tabular}
\end{table}

\noindent The only other non trivial formula VTS must traverse in $x$ is: $$x^2 < 1 \land \left( ( 3 < 2x \land -8x^2 + 12x < 5 ) \lor ( 3 \leq 2x \land -8x^2 + 12x - 5 = 0 ) \right)$$ with significant overlap in terms of test points, but also additionally identifying $x = \pm 1 + \epsilon$, making for a total of 11 unique non-degenerate test points in $x$. Again, there are fewer substitutions owing to exact roots than CAD, with VTS not identifying $x=\frac12$ or $x=\frac52$ as meaningful to substitute. Yet again all test points are substituted to receive \textbf{UNSAT}.

\subsubsection{CAC:}

Our implementation of the Cylindrical Algebraic Covering algorithm operates on this example as follows.

\begin{description}
	\item[$x=-1$:] Similarly yo the previous example, this would require $y^2 < 0$ and again we can not deduce anything around this value, as $-1$ is a root of the discriminant of $x^2+y^2-1$.
	\item[$x<-1$:] As in the first example we sample $x = -2$ and again find it to be unsatisfiable due to $y^2 < -3$, excluding the whole interval $(-\infty,-1)$.
	\item[$-1 < x$:] We have now excluded $(-\infty,-1]$ and sample $x=0$. While $y^2<1$ is still satisfiable, the second constraint evaluates to $(y-3/2)^2 < - \frac{5}{4}$ which is directly conflicting. The characterisation ends up excluding the interval $(-\infty, \frac{1}{2})$, essentially superseding all previous intervals.
	\item[$\frac{1}{2} < x$:] We sample $x=1$ and obtain a direct conflict with $y^2 < 0$. As with the first sample ($x=-1$) we only exclude the point interval $[1,1]$.
	\item[$\frac{1}{2} < x < 1$:] To take care of this interval we sample $x = \frac{3}{4}$ which finally needs both constraints to realize that no value for $y$ is feasible. The characterisation excludes exactly the interval $(\frac{1}{2},1)$. Note that the point $\frac{1}{2}$ remains uncovered.
	\item[$x=\frac{1}{2}$:] We now check the remaining point $x=\frac{1}{2}$ which directly conflicts with the second constraint. As $\frac{1}{2}$ is a root of the discriminant of the second constraints polynomial, only this point interval is excluded.
	\item[$1 < x$:] We continue with $x=2$, which is a direct conflict with the first constraint due to $y^2 < -3$, excluding $(1,\infty)$ and thereby completing the covering of the $x$-axis.
\end{description}
We have ascertained that no value for $x \in \R$ can satisfy the constraints and so may conclude UNSAT.

Note that, after pruning, the reasoning consists of the following components:
\begin{description}
	\item[1: $x \not\in (-\infty,\frac{1}{2})$] because of the second constraint,
	\item[2: $x \neq \frac{1}{2}$] because of the second constraint,
	\item[3: $x \not\in (\frac{1}{2},1)$] because of both constraints,
	\item[4: $x \neq 1$] because of the first constraint and
	\item[5: $x \not\in (1,\infty)$] because of the first constraint.
\end{description}

\subsubsection{Human:}
\font\manual=manfnt at 6pt
\def\X{\raise 4pt\hbox{{\manual\char127}}}

There is no trivial human proof this time like in Example 1.  If we proceed as there we find conditions on $x$ which are compatible and can only conclude that $x \in (\frac{1}{2}, 1)$.  Manipulation of these restrictions into the constraints would find a small range of potential $y$-axis over that $x$-interval where both constraints could possibly be satisfied. In other words there is no ``quick win''.
There are two obvious proof approaches. Both require to know that at one point, say $x=1$, the two constraints are not simultaneously satisfied. We use {\X} to indicate points where geometric reasoning would seem to be necessary.
\begin{enumerate}
\item The resultant of the two circles {\X} is $18x^2-27x+\frac{45}4$, whose roots are not real, and \emph{a fortiori} not in $[\frac32,1]$, so the two curves do not cross over $[\frac32,1]$, and hence there are no solutions here.
\item Let $\xi$ be a value in $[\frac32,1]$. Then we  need {\X} to have 
\begin{equation}\label{eq:1}
y<\strut_2\sqrt{1-\xi^2}.
\end{equation} 
Similarly we need {\X} to have 
\begin{equation}\label{eq:2}
y>\frac32+\strut_1\sqrt{1-(\xi-\frac32)^2}\ge0.
\end{equation} 
We can square (\ref{eq:2}): 
\[
y^2>\frac94 +3\strut_1\sqrt{1-(\xi-\frac32)^2}+\left(1-(\xi-\frac32)^2\right)
\]
since it is an inequality of positive numbers, which also implies we can square (\ref{eq:1}): $y^2<1-\xi^2$. So 
\[
1-\xi^2>\frac94 +3\strut_1\sqrt{1-(\xi-\frac32)^2}+\left(1-(\xi-\frac32)^2\right).
\]
Hence 
\[
0>\frac94 +3\strut_1\sqrt{1-(\xi-\frac32)^2}+3\xi-\frac94=3\left(\xi+\strut_1\sqrt{1-(\xi-\frac32)^2}\right).
\]
Then we have 
\[
-\strut_1\sqrt{1-(\xi-\frac32)^2}>\xi,
\]
again an inequality of positive numbers, and so either  $1-(\xi-\frac32)^2>\xi^2$, or $1-2\xi^2+3\xi-\frac94>0$. This is not true when $\xi=1$, and the roots of this quadratic (which is essentially the resultant) are not in $[\frac12,1]$ (in fact they are not real). Hence the inequality is nowhere true.
\end{enumerate}
The first approach requires less geometric reasoning, and furthermore that reasoning is essentially uniform --- ``curves can only cross at roots of the resultant''.  It is in fact very similar to line 3 of the CAC proof. Hence, at least in this case, CAC has essentially produced one of the possible proofs that a human would.

\section{Conclusion}

For general quantifier elimination (Problem \ref{Pr:1}), we have two standard implemented methods in the literature: CAD and VTS. Both require a completeness result to accept their results, which is currently beyond the reach of formal proof.

For the purely existential version (Problem \ref{Pr:2}), where SAT is easy to verify, but UNSAT is hard, we have a third method: CAC. At least in easy cases, its execution induces proofs much closer to the human proof.  The trace of the algorithm and its output seem to often allow for verifiable results without reliance on a verified completeness result for the entire algorithm.   We acknowledge that no verification has yet been conducted --- we publish this paper to highlight the opportunity to the verification community and encourage their input. Is it possible to regard CAC as a tactic that can guide an automatic theorem prover?


\begin{landscape}

\begin{table}[p]
\caption{CAD cells constructed for the example in Section \ref{Ex:1}\label{T:Ex1}}
\centering
\begin{tabular}{l||c|c|c|c|c|c|c|c|c|c|c}
\footnotesize
$x$:&$<-1$&$=-1$&$-1<x<1$&$=1$&$1<x<2$&$=2$&$2<x<3$&$=3$&$3<x<5$&$=5$&$>5$
\\ \hline
	$y$:&---&$<f_1$&$<\strut_1f_1$&$<f_1$&---&---&---&$<f_2$&$<\strut_1f_2$&$<f_2$&---
	\\
	$y$:&&$=f_1$&$=\strut_1f_1$&$=f_1$&&&&$=f_2$&$=\strut_1f_2$&$=f_2$
	\\
	$y$:&&$>f_1$&$\strut_1f_1<y<\strut_2f_1$&$>f_1$&&&&$>f_2$&$\strut_1f_2<y<\strut_2f_2$&$>f_2$
	\\
	$y$:&&&$=\strut_1f_1$&&&&&&$=\strut_2f_2$&
	\\
	$y$:&&&$>\strut_2f_1$&&&&&&$>\strut_2f_2$&
\\ \hline
\#&1&3&5&3&1&1&1&3&5&3&1
\end{tabular}
\\where $f_1=\RootOf(x^2+y^2-1,y)$, $f_2=\RootOf((x-4)^2+y^2-1,y)$,\\and $\strut_1f_i$ refers to an ordered root of the polynomials (Notation \ref{N1}).
\end{table}

\begin{table}[p]
\caption{Cells for the example in Section \ref{Ex:2}\label{T:Ex2}}
\centering
\begin{tabular}{l||c|c|c|c|c|c|c|c|c|}
$x$:&$<-1$&$=-1$&$-1<x<\frac12$				&$=\frac12$					&$\frac12<x<1$				&=1							&$1<x<\frac52$				&$=\frac52$	&$>\frac52$
\\ \hline
$y$:&---&$<f_1$	&$<\strut_1f_1$				&$<\strut_1f_1$				&$<\strut_1f_1$				&$<f_1$						&$<\strut_1f_2$				&$<f_2$ & ---\\
$y$:&	&$=f_1$	&$=\strut_1f_1$				&$=\strut_1f_1$				&$=\strut_1f_1$				&$=f_1$						&$=\strut_1f_2$ 			&$=f_2$ & \\
$y$:&	&$>f_1$	&$\strut_1f_1<y<\strut_2f_1$&$\strut_1f_1<y<\strut_2f_1$&$\strut_1f_1<y<\strut_2f_1$&$>f_1$						&$\strut_1f_2<y<\strut_2f_2$&$>f_2$ & \\
$y$:&	&		&$=\strut_2f_1$				&$=\strut_2f_1$				&$=\strut_2f_1$				&$f_1<y<\strut_1f_2$		&$\strut_1f_2<y<\strut_2f_2$&		& \\
$y$:&	&		&$>\strut_2f_1$				&$\strut_2f_1<y<f_2$		&$\strut_2f_1<y<\strut_1f_2$&$\strut_1f_2<y<\strut_2f_2$&$>\strut_2f_2$				&		& \\
$y$:&	&		&							&$=f_2$						&$=\strut_1f_2$				&$=\strut_2f_2$				&							&		& \\
$y$:&	&		&							&$>f_2$						&$\strut_1f_2<y<\strut_2f_2$&$>\strut_2f_2$				&							&		& \\
$y$:&	&		&							&							&$=\strut_2f_2$				&							&							&		& \\
$y$:&	&		&							&							&$>\strut_2f_2$				&							&							&		&
\\ \hline
\#	&1	&3		&5							&7							&9							&7							&5							&3		&1
\end{tabular}
\\where $f_1=\RootOf(x^2+y^2-1,y)$, $f_2=\RootOf\left((x-\frac32)^2+(y-\frac32)^2-1,y\right)$.
\end{table}

\end{landscape}

\end{document}